\documentclass[12pt,preprint]{aastex}
\shorttitle{Composition and Precursor for Crab Nebula}
\shortauthors{MacAlpine et al.} 
\received{2008 June}
\begin{document}

\title{THE CRAB NEBULA'S COMPOSITION\\ AND PRECURSOR STAR MASS}

\author{Gordon M. MacAlpine and Timothy J. Satterfield}
\affil{Department of Physics and Astronomy, Trinity University, San Antonio, TX 78212}

\begin{abstract}
We present results of new photoionization calculations for investigating
gaseous regions that represent potentially expected stages of nuclear
processing in the Crab Nebula supernova remnant.  In addition to gas resulting
from CNO-processing and oxygen-burning, as previously reported, 
a large component of the nebula appears to be carbon-rich. 
These results suggest that the precursor star had an initial mass $\gtrsim$~9.5~M$_{\odot}$. 
\end{abstract}
  
\keywords{ISM: individual (Crab Nebula)---nuclear reactions, nucleosynthesis, abundances--- 
supernovae: individual (SN 1054)---supernova remnants}

\section{INTRODUCTION}

The Crab Nebula (M1~=~NGC1952) is generally recognized as the remnant
of the core-collapse supernova SN1054. Estimates of the precursor star's initial mass have been 
in the range $8-12$~M$_{\odot}$ 
(see Davidson \& Fesen 1985 and references therein).  The visible remnant contains at least $1-2$~M$_{\odot}$  
of He-rich line-emitting gas (MacAlpine \& Uomoto 1991)\footnote{Fesen et al. 1997 used global line photometry
to suggest a higher nebular mass, approaching that once postulated by MacAlpine et al. 1989.  However, as 
discussed in MacAlpine \& Uomoto 1991, large spatial variations in the He/H ratio cause spuriously large mass 
estimates based on global line photometry.  More accurate emitting gas estimates should take into account
spatially resolved line photometry in order to allow for the effects of helium abundance variations on 
N(S)/N(H), which is used along with $[$S~II$]$ line intensities to estimate the amount of neutral gas present.}, 
the neutron star probably has about 1.4~M$_{\odot}$,
and the outer layers of the precursor may have left in a pre-supernova wind (Nomoto et al. 1982) 
or a shockwave (e.g., Chevalier 1977; Davidson \& Fesen 1985).

Knowledge of the chemical composition of the remnant is vitally important for a better understanding
of the precursor star's initial mass and associated details of the core-collapse event.  According to  
Nomoto et al. (1982), an $8-9.5$~M$_{\odot}$ star would likely have undergone an O-Ne-Mg core collapse 
following electron capture, whereas a star more massive than 9.5~M$_{\odot}$ would have developed an Fe-rich 
core.  A distinguishing factor between these two possible scenarios would be the nebular 
carbon abundance.  As pointed out by Nomoto (1985), for a precursor with initial mass less than about 9.5~M$_{\odot}$,
high N and low C mass fractions (compared with solar) would exist from CNO-cycle processing in the He-rich gas.  
On the other hand, for precursor mass $>$~9.5~M$_{\odot}$, helium-burning and nitrogen-processing via 
$^{14}$N($\alpha$,$\gamma$)$^{18}$F($\beta$$^{-}$$\nu$)$^{18}$O($\alpha$,$\gamma$)$^{22}$Ne ought to have 
produced high C and low N.  In addition, models for a roughly 10~M$_{\odot}$ precursor (Woosley et al. 1980; 
Woosley \& Weaver 1986) have suggested the existence of off-center neon and oxygen flashes just prior 
to the core collapse, whereas such events may not be expected in lower mass stars (Hillebrandt et al. 1984; 
Nomoto 1984).

The necessary chemical composition information should be obtainable.  Because of its young age 
and location roughly 180~pc from 
the Galactic plane, the Crab Nebula is relatively uncontaminated by interstellar material.   
In addition, its line-emitting gas is ionized and heated primarily by locally-generated synchrotron 
radiation (rather than shock heating), so the ionization, thermal, and 
chemical characteristics of the gas can be accurately analyzed using numerical photoionization codes.

Although there has been a long history of observational and theoretical investigations aimed at understanding 
abundances in the Crab Nebula, to some extent the results have been inconclusive or contradictory.  
As summarized by Henry (1986), Henry and MacAlpine (1982) and Pequignot and Dennefeld (1983) 
compared measured line-intensity data available at the time with photoionization calculations.
In both studies, the deduced C and N mass fractions were roughly solar or below, but there were problems
with interpretation of near-infrared $[$C~I$]$ emission lines (see below).
Davidson et al. (1982) and Blair et al. (1992) also obtained satellite observations of the ultraviolet 
C~IV and $[$C~III$]$ lines for certain locations, and they concluded no definitive evidence for a carbon excess,
although the latter study left open the possibility.  More recently, using extensive 
long-slit spectroscopy covering much of the nebula, MacAlpine et al. (1996) deduced high nitrogen 
mass fractions in some filamants and also high abundances for products of oxygen-burning, 
such as sulfur and argon, at other locations.  All of the above results, taken together, have generally been 
interpreted as suggestive of a precursor in the $8-9.5$~M$_{\odot}$ range, although high sulfur and argon abundances 
would appear to argue for a more massive star.

In this paper, we report on the results of new photoionization computations (summarized by MacAlpine et al. 2007b), 
which were developed using new line-intensity measurements 
which extend to 1~\micron\ (MacAlpine et al. 2007a, hereinafter Paper~1). These calculations suggest that  
strong $[$N~II$]$~$\lambda$$\lambda$6548,6583 emission lines measured at many locations by MacAlpine et al.
(1996, 2007a) do {\it not} necessarily imply high nitrogen abundances, while strong measured $[$C~I$]$~$\lambda$9850 
emission {\it can} be interpreted as indicative of high carbon mass fractions.  In addition, high mass fractions 
for S and Ar in some locations are confirmed.  These results argue {\it consistently} for a precursor star 
initial mass $\gtrsim$~9.5~M$_{\odot}$.

\section{PHOTOIONIZATION COMPUTATIONS AND NUCLEAR PROCESSING REGIONS}

Photoionization models were produced using the numerical code Cloudy, version 06.02 
(described in Ferland, G. J. et al. 1998).  Guided by line-intensity measurements and correlations in Paper~1, 
we developed plane-parallel computations for a range of physical conditions and chemical abundances 
in efforts to investigate gaseous regions that may represent expected stages of nuclear 
processing for relevant stellar models (e.g., Woosley \& Weaver 1995).

At least three nuclear processing domains may be inferred from the observations in Paper~1.
Figure~1 of this paper (reproduced from Figure~4 of Paper~1) shows the correlation between 
measured, H$\alpha$-normalized, reddening-corrected $[$N~II$]$$\lambda$6583 and $[$S~II$]$$\lambda$6731 
line intensities for 180 widely-distributed nebular locations.  It was suggested in Paper~1 
that the strongest nitrogen emission may represent gas which has progressed
no further than the CNO-cycle, whereas the apparent majority of measurements (e.g., with normalized $[$N~II$]$$\lambda$6583 
over the range 1 to 2.5) could represent material in which some helium-burning has taken place and nitrogen has been
converted into neon.  The strongest sulfur emission was postulated to represent more advanced processing
through oxygen-burning. In this paper, these regions of Figure~1 will be called nuclear processing ``Domains 1, 2, and 3,''
respectively. 

\subsection{DOMAIN~2 (HELIUM-BURNING AND NITROGEN DEPLETION)}

We begin by discussing details of our analysis for Domain~2 because it involves the most prevalent nebular gas.
In addition, the relevant photoionization models also help to understand the other domains. 

First, we developed a representative ``median observed spectrum'' for Domain~2.  This involved a subset of 19 spectra 
for which we have both optical and near-infrared measurements (see Paper~1) and for which
the H$\alpha$-normalized $[$N~II$]$$\lambda$6583 emission lies in the range 1--2.5.  For each emission-line measurement,
we identified a median value and combined those into a ``median spectrum,'' as given in Table~1.
Then we adjusted photoionization model input parameters and chemical abundances in order to obtain
a satisfactory fit to the observed data.  

Following considerable experimentation, all models presented here employ the ionizing synchrotron radiation 
spectrum given by Davidson \& Fesen (1985), with added points across the unobserved ultraviolet range
as suggested in the documentation which accompanies the Cloudy code.
In addition, we employed a constant hydrogen density of 3000~cm$^{-3}$
and log ionization parameter~=~-3.2. The ionization parameter is defined as 
U~=~$\Phi$/cN, where $\Phi$ is the flux of photons more energetic than 1~Rydberg striking the face of a cloud,
c is the speed of light, and N is the total density of hydrogen.  The ionizing spectrum did a good job of
producing typically observed HeII/HeI line ratios (e.g., MacAlpine et al. 1989), the adopted
hydrogen density led to calculated electron densities in S$^+$ zones comparable with those measured
by Fesen \& Kirshner (1982), and the log~U was necessary to obtain a consistent overall
emission line spectrum that matches the observations, in particular the 
$[$S~III$]$$\lambda$9069/$[$S~II$]$$\lambda$6731 ratio.  

The ionization parameter employed here is almost a factor of 10 lower than that which is necessary
to account for far-ultraviolet C~IV and $[$C~III$]$ lines (see Blair~et~al.~1992).  Henry and MacAlpine (1982)
noted similar ionization parameter differences when comparing photoionization calculations that
could match optical and ultraviolet observations separately. We do not consider this to be a problem.  
As shown convincingly by Blair et al., much of the
ultraviolet line emission probably comes from lower-density, more diffuse gas, compared
with the gas in optical line-emitting filaments being analyzed here.

For our Domain~2 model, the helium abundance was set at 89\%
by mass fraction, consistent with some stellar model expectations (Woosley \& Weaver 1995)
for a region with partial helium-burning and also with numerous observations of He~I$\lambda$5876 emission 
(e.g., Uomoto \& MacAlpine 1987; MacAlpine et al. 1989; Paper~1). 

Figure~2 presents the computed Domain~2 hydrogen and helium ionization fractions as a function of depth into
an emitting gas cloud, with ionizing radiation entering from the left.  We note that the He$^{+2}$
and He$^{+}$ zones do not extend far into the cloud.  Because of the high helium abundance,
ionizing photons more energetic than 54.4~ev (the ionization potential of He$^{+}$)
and 24.6~ev (the ionization potential of He$^{0}$) are absorbed before they can penetrate
a substantial distance.  The hydrogen ionized depth is about 10$^{16}$~cm, which is consistent
with observations for Crab Nebula filaments (see Davidson \& Fesen 1985).

Figure~3 shows the computed electron temperature and density, plotted as a function of distance into the cloud.
The electron density is roughly 8000~cm$^{-3}$ where helium is ionized, drops to 3000~cm$^{-3}$ through
the H$^{+}$ zone, and then falls to 400-800~cm$^{-3}$.  The temperature peaks near
1.2$\times$10$^{4}$~K, is about 6-8$\times$10$^{3}$~K through the H$^{+}$ zone, and then remains
above about 6000~K to the depth plotted.

Figure~4 gives the modeled carbon and nitrogen ionization.  Because the C$^{+}$
ionization potential is comparable with that for He$^{+}$, and the C$^{0}$ ionization potential
is below that for H$^{0}$, the dominant ionization stage for carbon is C$^{+}$,
all the way from the He$^{+}$ zone to where the calculations were stopped at a depth of approximately seven times
the hydrogen ionization zone thickness.
In addition, because the N$^{+}$ and N$^{0}$ ionization potentials are slightly above (within 5~ev of)
those for He$^{0}$ and H$^{0}$, respectively, N$^{+}$ occupies an extensive, very proficient
``$[$N~II$]$ emitting zone,'' in which the temperature and electron density effectively support collisional excitation.

Figure~5 presents an expanded view of the lower 10\% of the vertical scale from Figure~4.
This illustrates how a relatively small, but significant amount of C$^{0}$ exists in the zones
where helium and hydrogen are ionized.  Progressing from left to right in the figure, 
there is a small peak in the C$^{0}$ fraction where the electron density is highest due to ionized helium.  
Then there is a gradual C$^{0}$ increase as the temperature declines, and finally there is essentially 
no C$^{0}$ beyond the hydrogen ionization edge where the electron density becomes very low.  
These distribution characteristics for C$^{0}$ are consistent what would be expected from
C$^{+}$$\rightarrow$C$^{0}$ recombination.

Figure~6 shows selected, computed line emissivities for this model, as a function of depth into the cloud.
First, we note that $[$N~II$]$~$\lambda$6584 is ``off the chart'' because of electron collisional
excitation in the warm, high-electron-density $[$N~II$]$ emitting zone. 
H~I~$\lambda$6563 illustrates what would be expected from recombination, and $[$C~I$]$~$\lambda$9850
shows a similar distribution.  As discussed above, the latter results from C$^{+}$$\rightarrow$C$^{0}$ recombination
followed by electron collisional excitation.  The $[$O~I$]$~$\lambda$6300 line arises from collisional
excitation of O$^{0}$, and it is primarily important in the neutral hydrogen region because of 
charge exchange processes between hydrogen and oxygen.

Previously, high nitrogen mass fractions for many locations were deduced from observations of strong
$[$N~II$]$~$\lambda$$\lambda$6548,6583 emission and models using the Cloudy code in MacAlpine et al. (1996).
However, those models did not adequately take into account very high helium abundance
and its effect on broadening the $[$N~II$]$ emitting zone.
The helium mass fraction over most of the nebula ranges from about 75\% to greater than 90\% (Uomoto \& MacAlpine 1987).
For such helium-rich regions, strong $[$N~II$]$ lines
would {\it not} necessarily imply nitrogen mass fractions greater than solar.

It has been known for decades that $[$C~I$]$~$\lambda$9850 emission is very strong in parts
of the Crab Nebula (Dennefeld \& Andrillat 1981; Henry et al. 1984).
Pequignot \& Dennefeld (1983) tried to match the observed strengths of this line by postulating
a high dielectronic recombination rate, and Henry et al. (1984) postulated that collisional excitation
by H$^{0}$ could play an important role.
Neither group was successful in using $[$C~I$]$~$\lambda$9850 for reliable carbon abundance estimates.
Therefore, difficult measurements
of the ultraviolet $[$C~III$]$ and C~IV emission were attempted as noted previously.
However, production of those lines may not be well understood either, in the sense that model
predictions can be wrong by as much as a factor of two depending on how He~II~L$\alpha$ absorption
is treated (Eastman et al. 1985).  In the present calculations, we understand the production of 
$[$C~I$]$~$\lambda$9850 emission as being due to thermal electron collisional excitation.  
This $^{1}$D$_{2}$$\rightarrow$$^{3}$P$_{2}$ transition is an analog of the $[$O~III$]$~$\lambda$5007 line.  
Making use of effective collision strengths
given by Pequignot \& Aldrovandi (1976), we manually calculated emissivities at various locations
and matched the model-computed emissivities exactly.

Predicted relative emission-line intensities from this model are shown in Table~1, where it may be seen
that there is a reasonable match with the Domain~2 median spectrum for element mass fractions 
given in Table~2.  The deduced Domain~2 nitrogen mass fraction is less than half of its solar value.
We also note that the carbon mass fraction required to produce the observed median $[$C~I$]$$\lambda$9850 emission 
in Domain~2 is roughly 6 times its solar value, and the deduced oxygen mass fraction is very close to solar.  
A surprising result is the sulfur mass fraction necessary to acount for the observed $[$S~II$]$$\lambda$6731
and $[$S~III$]$$\lambda$9069 emission, which is less than a third of its solar value.  

\subsection{DOMAIN~1 (STRONGEST NITROGEN EMISSION)}

The photoionization calculations for Domain~1 gas employed the same ionizing spectrum, hydrogen density, 
and ionization parameter as discussed above; and the model characteristics are similar to those 
in Figures~2-6.  The median measured spectrum and computed line intensities are shown in Table~1.
The measured, normalized $[$N~II$]$$\lambda$6583 values for Domain~1
range from 3.1 to 5.6, and the adopted median is 4.5.  Unfortunately, the subset 
of near-infrared data with measurable $[$C~I$]$$\lambda$9850 did not include any nebular locations
with $[$N~II$]$$\lambda$6583~$>$~2.5 (see Figure~10 of Paper~1). Therefore, no $[$S~III$]$ or $[$C~I$]$
line values are given for this domain in Table~1.   

As may be seen in Table~2, the helium mass fraction required to match the observed He~1$\lambda$5876 line
is again 89$\%$ for Domain~1, which is consistent with expectations from stellar models in a zone
containing largely CNO-processed material. However, we note that the inferred nitrogen mass fraction is only solar,
because $[$N~II$]$$\lambda$6583 arises so effectively in the $[$N~II$]$ emitting zone discussed above
and illustrated in Figures~4 and 6.  This implies some gas mixing or helium-burning.  The deduced 
oxygen mass fraction is solar, while the data indicate that
sulfur is significantly less than solar, as in Domain~2.  The carbon mass fraction input 
to the Domain~1 model was {\it assumed} to be 0.5~solar. A lower-than-solar carbon mass fraction
would be consistent for a region in which CNO-processing has taken place.

\subsection{DOMAIN~3 (STRONGEST SULFUR EMISSION)}

The measured spectrum and computed line intensities for Domain~3 are also shown in Table~1.
By combining optical data for locations with measured $[$S~II$]$$\lambda$6731 over the range 3-5.9
(see Figure~1) together with a subset of near-infared data, we derived a representative median
spectrum for this domain.  The best-fit photoionization model was similar to that for Domains~1 and 2, 
except it was necessary in this case to lower log~U to -3.5 to match the 
$[$S~III$]$$\lambda$9069/$[$S~II$]$$\lambda$6731 ratio.

As seen in Table~2, the adopted helium mass fraction to match observed median He~I~$\lambda$5876 
is still 89\% for Domain~3, which would seem to imply gas mixing for regions
in which significant processing beyond helium-burning has taken place.  As expected, the nitrogen mass fraction
is now well below solar.  The deduced carbon and oxygen mass fractions are both about 10 times solar.
We note that the $[$O~I$]$~$\lambda$6300 line emissivity plotted in Figure~6 illustrates that it arises
in a different region from $[$C~I$]$~$\lambda$9850, so the inferred, correlated carbon and oxygen mass fractions
are not the result of high-optical-depth regions in the emitting gas.  This also supports the idea that the
$[$C~I$]$~$\lambda$9850:$[$O~I$]$~$\lambda$6300 line correlations in Figure~9 of Paper~1
result from carbon and oxygen abundances having been produced together by nuclear processing.

The sulfur mass fraction is roughly 4 times solar in this domain where it is hypothesized that
oxygen-burning has taken place.  Further support for oxygen-burning comes from the high inferred argon
mass fraction in this domain, as well as Figure~6 of Paper~1, 
which shows a very tight correlation 
between $[$Ar~III$]$ and $[$S~II$]$ emission, supporting the idea that sulfur and argon are produced
together by the same process.  
Finally, a correlation shown in Figure~8 of Paper~1 is consistent with the hypothesis that iron-peak nuclei
like nickel are produced by ``alpha-rich freeze out'' (see Woosley \& Weaver 1995; Jordan et al. 2003)
in Domain~3 regions containing enhanced silicon-group elements from oxygen-burning.

\section {SUMMARY AND DISCUSSION}

The photoionization calculations presented here, compared with emission-line measurements
from Paper~1, are suggestive of three nuclear processing ``domains'' for gas in the Crab Nebula.

1. Gas with the strongest $[$N~II$]$ lines has roughly solar nitrogen and oxygen mass fractions, probably a carbon 
mass fraction less than solar, and significantly less-than-solar sulfur and argon mass fractions.  
This domain with 89\% helium may be largely CNO-cycle processed gas in which modest helium-burning and
nitrogen depletion have taken place.

2. The majority of the emitting nebular gas, with $[$N~II$]$$\lambda$6583/H$\alpha$ in the range 1--2.5, also has
$\sim$~89\% helium implied by its median observed spectrum. For this domain, the deduced nitrogen 
mass fraction is depleted to less than half its solar value.  In addition, the carbon
mass fraction is about 6 times solar, while oxygen is solar, and sulfur and argon
have approximately a third to half of their solar values.  It is postulated that this domain
represents gas in which significant helium-burning has taken place.

3. The third domain examined here has the highest $[$S~II$]$$\lambda$6731 measured line intensities.
We find that the inferred helium mass fraction is still comparable with the other domains,
and the nitrogen mass fraction is about a fourth of solar.  The carbon and oxygen mass fractions
are roughly 10 times solar, while the sulfur and argon mass fractions are about 4-6 times solar.
This is considered to be a domain in which oxygen-burning has occurred and some mixing has taken place.
  
Gas resulting from the CNO-cycle and gas with products of oxygen-burning
were noted previously (MacAlpine et al. 1996), where the former was assumed to be the primary
component due to strong observed nitrogen emission throughout most of the nebula.
However, the current computations indicate that strong $[$N~II$]$$\lambda$$\lambda$6548,6583
emission does {\it not} necessarily imply high nitrogen abundance,
whereas strong $[$C~I$]$$\lambda$$\lambda$9823,9850 emission from much of the gas {\it does} 
imply high carbon abundance.

The finding of high carbon mass fraction throughout much of the Crab Nebula gas is particularly
significant because it suggests that the supernova precursor star had an initial mass $\gtrsim$~9.5~M$_{\odot}$.
However, it is important to consider whether the near-infrared emission-line measurements
presented in Paper~1, and used here, may be biased toward high $[$C~I$]$$\lambda$9850 emission.
As discussed in Paper~1, the near-infrared data were obtained during non-optimal seeing, and 
$[$C~I$]$ measurements were made only where potentially useful $[$S~III$]$ or $[$C~I$]$ line-emission 
could be identified in the two-dimensional, near-infrared spectral images. To investigate the possibility
of consequent biasing, we compared the current measurements with $[$C~I$]$$\lambda$$\lambda$9823,9850 
emission reported by Henry et al. (1984).  
The ranges of measured $\lambda$9850 from that study and this one are comparable, with marginally 
lower and higher values in the larger data set used here.  However, we still suspect that the current 
median spectrum for Domain~2 may be biased toward stronger $[$C~I$]$ in view of how 
the measured positions were selected.  That having been said, we note that the {\it primary} reason 
for the current finding of high carbon mass fractions is a photoionization calculation whereby we can derive potentially 
useful carbon abundances from $[$C~I{$]$ emission.  Furthermore, even though our $\lambda$9850 {\it median}
values may bias the inferred carbon mass fraction for a domain, the individual measurements
do imply that carbon-rich gas exists at numerous loctions in the nebula.  In any case, a more extensive
set of near-infrared spectra would be very useful for understanding the overall nebular carbon.

We should also reflect here on the fact that the photoionization code gave results for plane-parallel 
geometrical situations, whereas line-emitting gas in the Crab Nebula may be immersed in ionizing 
synchrotron radiation from all directions.  Therefore, we developed algorithms for converting the 
plane-parallel results to what would be expected for convex cylindrical and spherical clouds.  
Potentially noteworthy changes in H$\alpha$-normalized lines occurred for only $[$S~II$]$$\lambda$6731
and $[$O~I$]$$\lambda$6300 emission.  In going from plane-parallel to spherical geometry, the computed
$\lambda$6731/H$\alpha$ could be reduced by as much as 30\%, and $\lambda$6300/H$\alpha$ 
could be reduced by about 40\%.  

Because adopted ionization parameters for the photoionization calculations
involved matching the observed $[$S~III$]$$\lambda$9069/$[$S~II$]$$\lambda$6731 ratio, an employed
ionization parameter could be slightly lower for a convex spherical geometry.  However, 
we investigated this possibility and found that it would probably not
have a significant effect on the conclusions presented here.  We also considered whether 
convex spherical or cylindrical clouds might affect our deduced low sulfur mass fractions for Domains~1 and 2.
By using a spherical gometry, we could increase a sulfur mass fraction by perhaps 30\%,
but it (and that for argon) would still be low in these domains.  Finally, we note that Henry \& MacAlpine (1982)
also reported lower-than-solar sulfur for particular filaments.  We believe the low sulfur mass fractions 
in Domains~1 and 2 to be valid results of this investigation.

\acknowledgments

This work was supported by Trinity University and by an endowment from the late
Mr. Gilbert Denman.

\clearpage

\begin{deluxetable}{lrrrrrr}
\tablewidth{0pt}
\tablecaption{Measured And Computed Line Intensities\tablenotemark{a}
\label{tab1}}
\tablecolumns{7}
\tablehead{
\colhead{} &\multicolumn{2}{c}{Domain 1} &\multicolumn{2}{c}{Domain 2} &\multicolumn{2}{c}{Domain 3} \\
\colhead{Line} &\colhead{Measured} &\colhead{Computed} &\colhead{Measured} &\colhead{Computed} &\colhead{Measured} &\colhead{Computed}
}
\startdata
He~I~$\lambda$5876 & 0.26 & 0.27 & 0.28 & 0.26 & 0.28 & 0.25 \\
$[$O~I$]$~$\lambda$6300 & 0.40 & 0.41 & 0.32 & 0.34 & 0.58 & 0.44 \\
$[$N~II$]$~$\lambda$6583 & 4.5 & 4.8 & 1.6 & 1.9 & 0.50 & 0.52 \\
$[$S~II$]$~$\lambda$6731 & 0.64 & 0.66 & 0.85 & 0.82 & 4.0 & 3.6 \\
$[$Ar~III$]$~$\lambda$7136 & 0.05 & 0.05 & 0.11 & 0.10 & 0.45 & 0.35 \\
$[$S~III$]$~$\lambda$9069 & --- & 0.22 & 0.26 & 0.27 & 0.83 & 0.90 \\
$[$C~I$]$~$\lambda$9850 & --- & 0.04 & 0.45 & 0.45 & 0.74 & 0.70 \\

\enddata
\tablenotetext{a}{Normalized to H$\alpha$.}

\end{deluxetable}

\clearpage

\begin{deluxetable}{lrrr}
\tablewidth{0pt}
\tablecaption{Deduced Element Mass Fractions
\label{tab2}}
\tablecolumns{4}
\tablehead{
\colhead{} &\colhead{Domain 1} &\colhead{Domain 2} &\colhead{Domain 3} \\
\colhead{Element} &\colhead{Mass Fraction} &\colhead{Mass Fraction} &\colhead{Mass Fraction} 
}
\startdata
Helium & 89$\%$ & 89$\%$ & 89$\%$ \\
Nitrogen & solar & 0.35 solar & 0.28 solar \\
Carbon & 0.50 solar & 6 $\times$ solar & 11 $\times$ solar \\
Oxygen & solar & solar & 10 $\times$ solar \\
Sulfur & 0.21 solar & 0.28 solar & 4.2 $\times$ solar \\
Argon & 0.24 solar & 0.55 solar & 6.0 $\times$ solar \\
\enddata

\end{deluxetable}

\clearpage

\begin{figure}
\epsscale{1.0}
\plotone{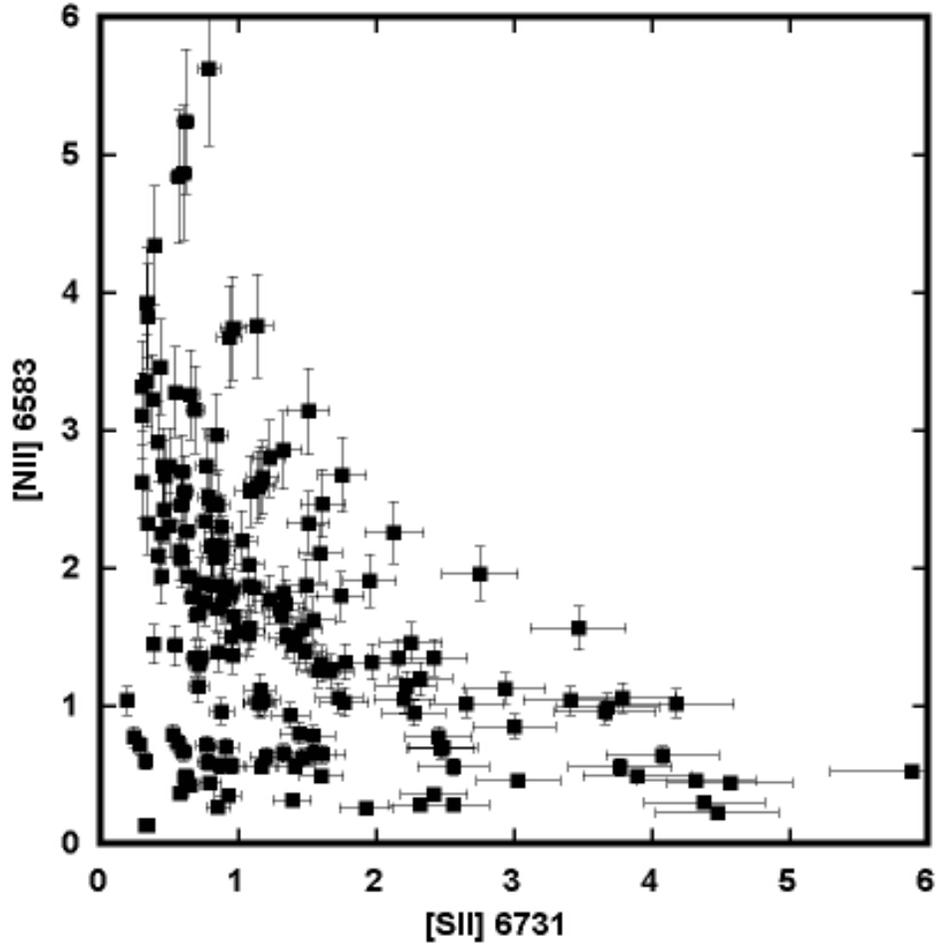}
\caption{Correlation between H$\alpha$-normalized $[$N~II$]$$\lambda$6583 and $[$S~II$]$$\lambda$6731 
measured line intensities from Figure~4 of Paper~1.
\label{fig1}}
\end{figure}
\clearpage

\begin{figure}
\epsscale{1.0}
\plotone{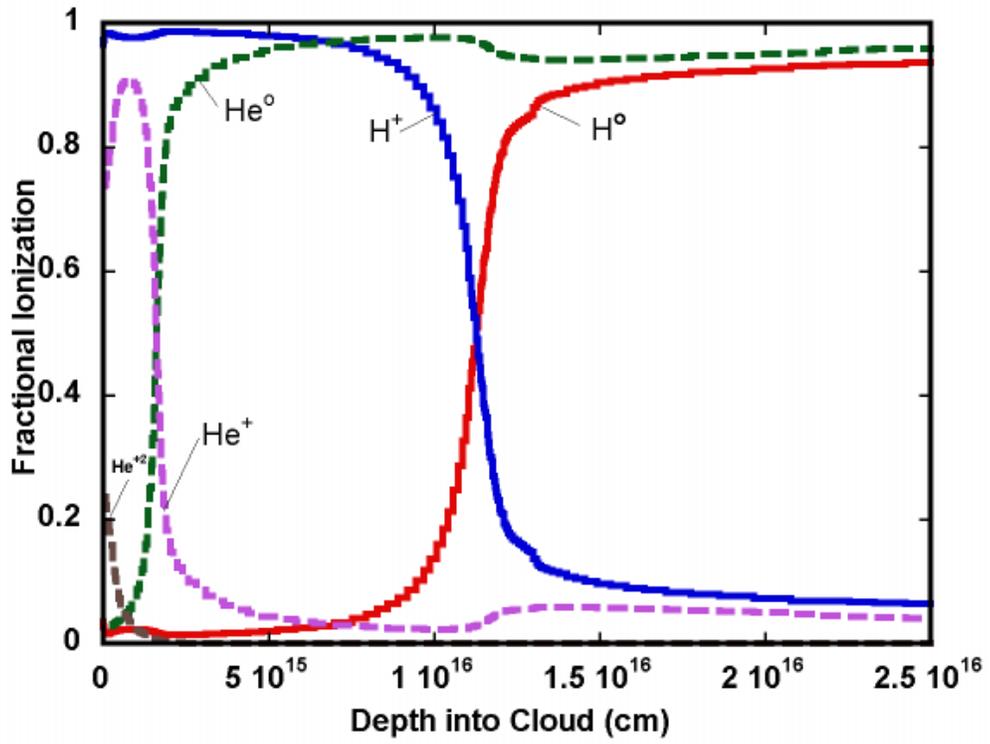}
\caption{Domain~2 hydrogen and helium ionization fractions as a function of distance into a cloud.
\label{fig2}}
\end{figure}
\clearpage

\begin{figure}
\epsscale{1.0}
\plotone{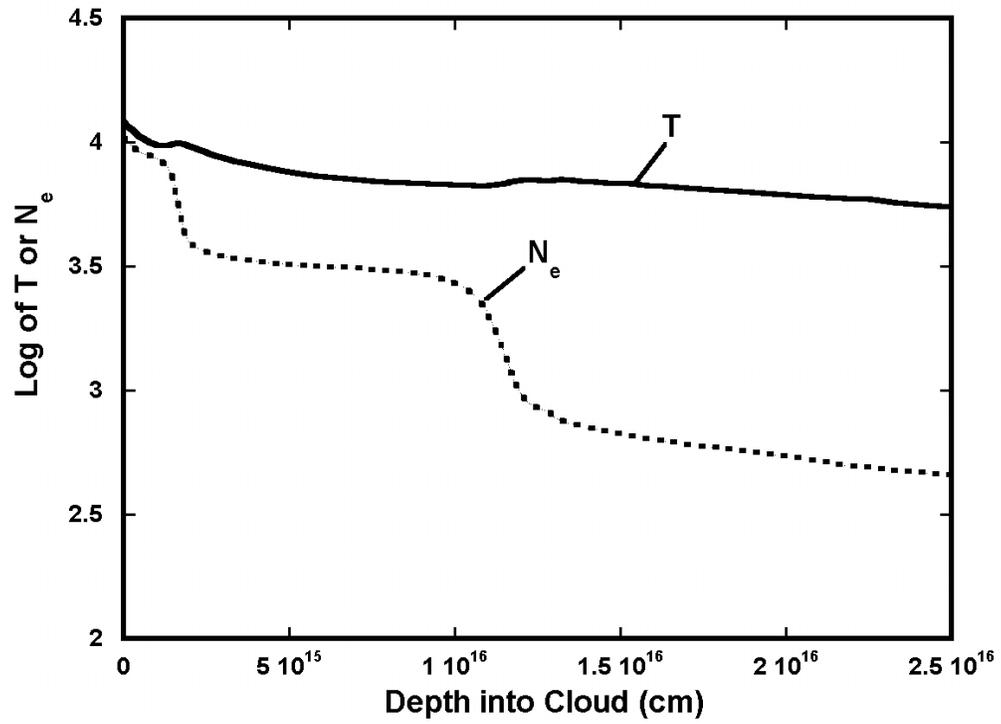}
\caption{Domain~2 temperature and electron density as a function of distance into a cloud.
\label{fig3}}
\end{figure}
\clearpage

\begin{figure}
\epsscale{1.0}
\plotone{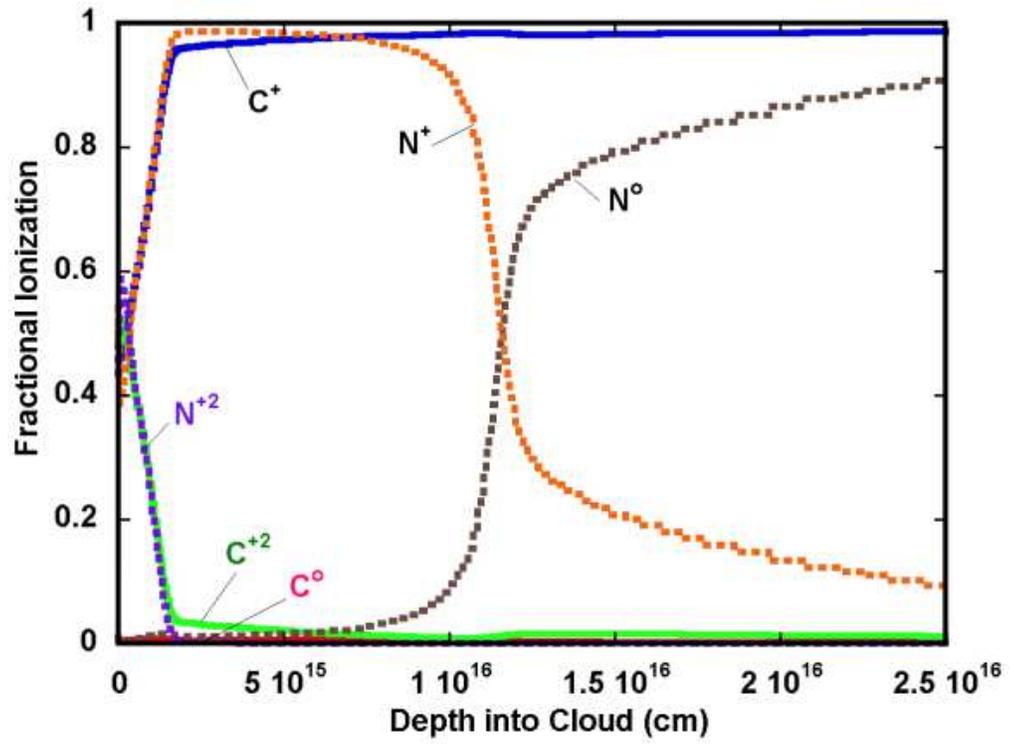}
\caption{Domain~2 carbon and nitrogen ionization fractions as a function of distance into a cloud.
\label{fig4}}
\end{figure}
\clearpage

\begin{figure}
\epsscale{1.0}
\plotone{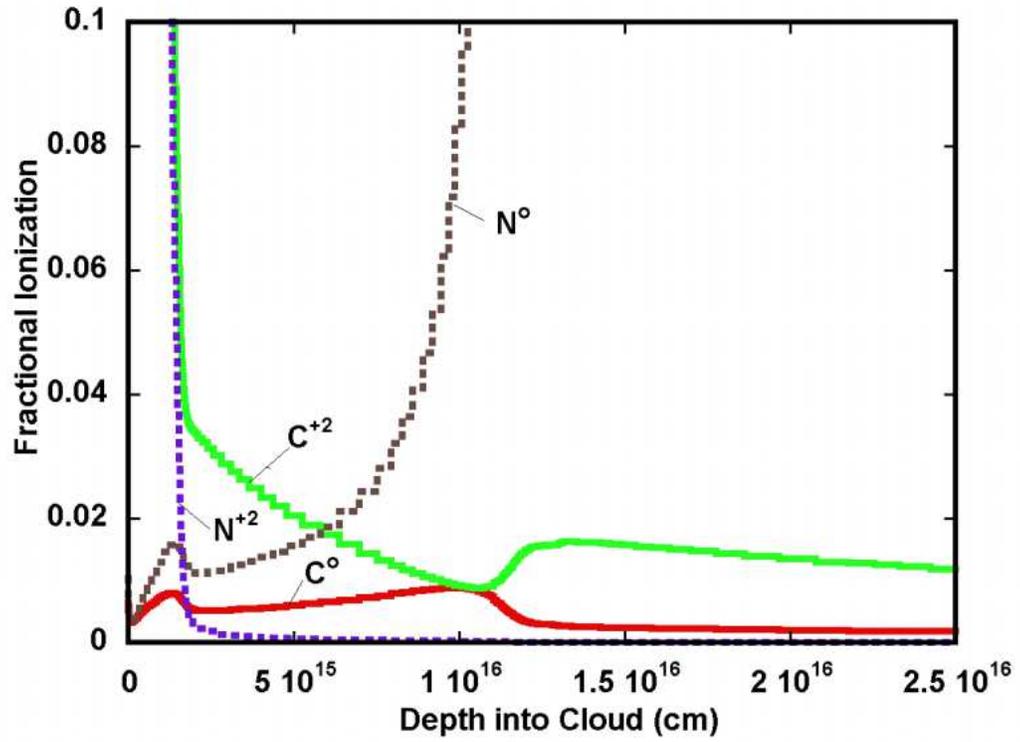}
\caption{Expanded view of carbon and nitrogen ionization fractions below the value 0.1 on the vertical axis of 
Figure 4.
\label{fig5}}
\end{figure}
\clearpage

\begin{figure}
\epsscale{1.0}
\plotone{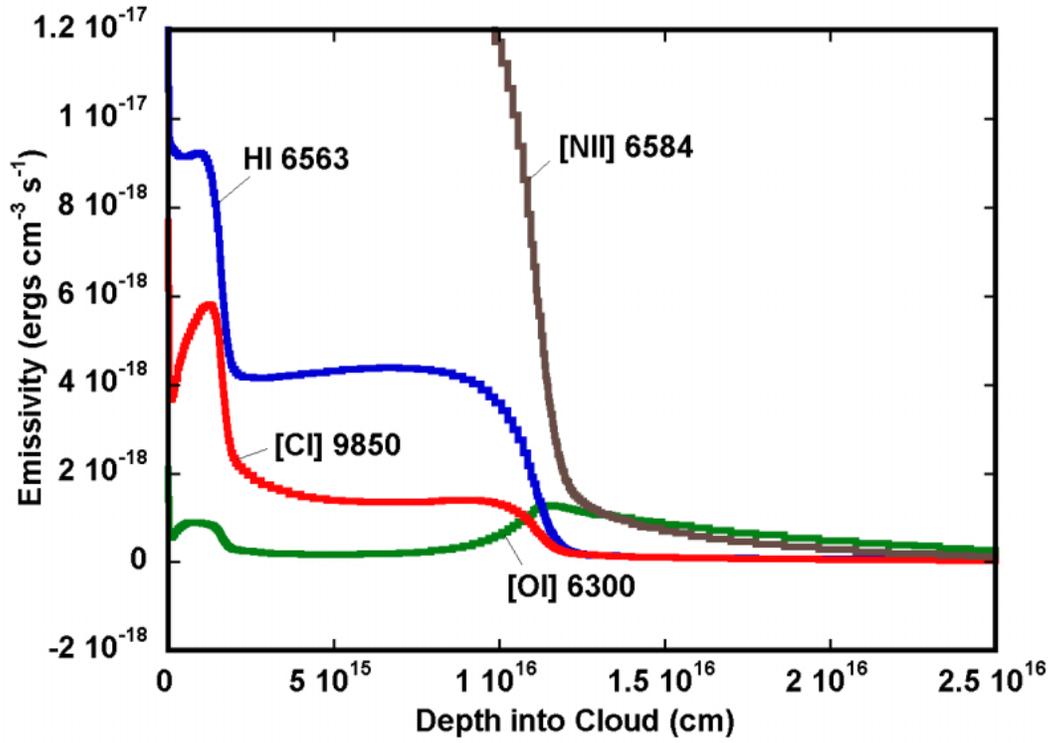}
\caption{Selected Domain~2 line emissivities as a function of distance into a cloud. 
\label{fig6}}
\end{figure}
\clearpage

\end{document}